# Large room temperature spin-to-charge conversion signals in a few-layer graphene/Pt lateral heterostructure

Wenjing Yan[1,‡], Edurne Sagasta[1,‡], Mário Ribeiro[1], Yasuhiro Niimi[2], Luis E. Hueso[1,3,*] & Fèlix Casanova[1,3,*]

[1]CIC nanoGUNE, 20018 Donostia-San Sebastian, Basque Country, Spain.
[2]Department of Physics, Graduate School of Science, Osaka University, 1-1 Machikaneyama, Toyonaka, Osaka 560-0043, Japan
[3]IKERBASQUE, Basque Foundation for Science, 48011 Bilbao, Basque Country, Spain.

[‡]These authors contributed equally to this work
*E-mail: l.hueso@nanogune.eu; f.casanova@nanogune.eu

Electrical generation and detection of pure spin currents without the need of magnetic materials are key elements for the realization of full electrically controlled spintronic devices. In this framework, achieving a large spin-to-charge conversion signal is crucial, since considerable outputs are needed for plausible applications. Unfortunately, the values obtained so far have been rather low. Here we exploit the spin Hall effect by using Pt, a non-magnetic metal with strong spin-orbit coupling, to generate and detect pure spin currents in a few-layer graphene channel. Furthermore, the outstanding properties of graphene, with long distance spin transport and higher electrical resistivity than metals, allows us to achieve in our graphene/Pt lateral heterostructures the largest spin-to-charge voltage signal at room temperature reported so far in the literature. Our approach opens up exciting opportunities towards the implementation of spin-orbit-based logic circuits and all electrical control of spin information without magnetic field.

**Introduction**

A spintronic device with complete electrical functionality is attractive for its incorporation into the current charge-based integrated circuits. While advances have been made in the electrical control of spin transport[1-4], new approaches that allow electrical generation or detection of pure spin currents without using ferromagnetic materials (FM) as the spin source are also being developed[5-8]. In particular, in the emerging field of spin orbitronics[9,10], by uniquely exploiting the spin-orbit coupling (SOC) in non-magnetic materials, spin-to-charge current conversions have been realized using the spin Hall effect (SHE)[11-13], the Rashba-Edelstein effect[14,15], or the spin-momentum locking in topological insulators[16,17]. Magnetization switching of FM elements for memories[5-7] or the recent proposal of a scalable charge-mediated non-volatile spintronic logic[18] are applications based on SOC which can have a strong technological impact.

In the SHE, a pure spin current is generated from a charge current due to the strong SOC in a non-magnetic material. Reciprocally, the inverse spin Hall effect (ISHE) can be used for spin detection, since a spin current is turned into a measureable charge current[19]. The efficiency of this spin-to-charge interconversion is given by the spin Hall angle ($\theta_{SH}$). Since technological applications require large conversions, finding routes to maximize $\theta_{SH}$ has become a demanding task. The materials with highest reported yields are heavy metals with strong SOC, such as Pt[20,21], Ta[5] and W[22,23]. Recently, a clear route to enhance $\theta_{SH}$ of Pt has been unveiled[21]. Another path that has been explored to maximize the voltage output is the use of higher resistive spin Hall materials, although the potential enhancement is counteracted by the increased shunting of the induced charge current in the less resistive spin transport material, usually Cu or Ag[8]. It would be desirable to employ these spin-to-charge conversions to efficiently inject and detect pure spin currents in good spin transport materials with higher charge resistance.

Graphene, due to its weak spin-orbit coupling[24], is theoretically predicted and experimentally demonstrated to be an excellent material for spin transport[25-40]. Spin diffusion lengths of up to 90 μm have been reported at room temperature[31,32], a parameter found to be fairly insensitive to temperature[37-39]. However, electrical spin injection from FM sources suffers from a lack of reproducibility due to the required interfacial barriers[25,35,40], which are usually made of oxides that grow in an island mode on the graphene surface. An efficient and reliable spin injection and detection into such an outstanding spin transport material at room temperature using non-magnetic electrodes is of both fundamental and technological interests.

In our work, we demonstrate the generation (detection) of pure spin currents in a graphene-based lateral heterostructure by employing the SHE (ISHE) of Pt, avoiding the use of a FM source. Moreover, the large charge resistance of graphene as compared to the standard spin transport metals such as Cu and Ag eliminates completely the shunting effect, generating large output voltages. The spin-to-charge conversion signal in a graphene/Pt lateral device at room temperature is two orders of magnitude larger than the best performing ones previously reported that use metallic channels. Our concept of using charge-to-spin conversion to inject spin currents and spin-to-charge conversion to detect spin currents in graphene-based devices could open future applications of all electrical control of spin information without magnetic field.

**Results**
**Device structure.** We used the spin absorption method in lateral spin valve (LSV) devices to demonstrate spin generation and detection in graphene via the SHE and ISHE of Pt, respectively (see sketches in Fig. 1a). A SEM image of a complete device is shown in Fig. 1b. It consists of a 250-nm-wide flake of trilayer graphene (with a sheet resistance $R_{Gr}^{\blacksquare}$ = 1085 Ω and a carrier density n~$8 \times 10^{11}$ cm$^{-2}$) obtained via exfoliation[41], where spins are to be transported. Several ferromagnetic Co electrodes with their respective TiO$_2$ interfacial barriers are placed on top of the flake. The presence of TiO$_2$ between the Co electrode and the graphene channel leads to interface resistances between 10 and 42 kΩ. Additionally, Pt wires are placed on top of the graphene channel. We use very resistive Pt with $\rho_{Pt}$ = 99 (134) μΩcm at 50 (300) K and, therefore, a large spin Hall angle of $\theta_{SH}$ = 17.8 ± 2.0 (23.4 ± 2.5)% (ref. 21). Transport measurements are performed in a liquid-He cryostat with a superconducting magnet using a DC reversal technique[42-44]. See Methods for details on the device fabrication and measurements.

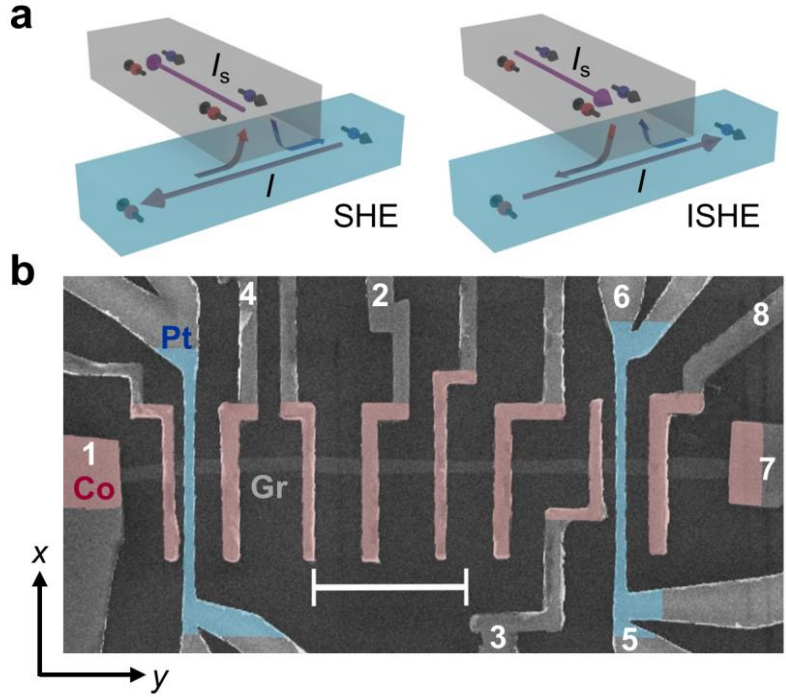

**Figure 1 | Illustration of the spin absorption method and scanning electron microscope (SEM) image of the device.** (a) Sketches of spin injection (left) and detection (right) using the SHE and ISHE of Pt, respectively, with the spin absorption technique, in which a pure spin current $I_S$ is vertically transferred between a non-magnetic spin transport channel (grey) and a metal with strong SOC (blue). (b) SEM image of a graphene-based spintronic device. It consists of standard LSVs with ferromagnetic Co electrodes with $TiO_2$ barrier (red color) placed adjacent to each other. This electrode configuration allows the study of the spin transport properties of graphene (grey color) using standard FM electrodes (see Fig. 2a for details). Additionally, Pt wires (blue color) are placed in between two pairs of Co electrodes. This extra configuration allows the study of spin absorption by Pt (see Fig. 3a for details) and spin current injection (via SHE) and detection (via ISHE) using Pt (see Fig. 4a for details). Scale bar is 3 μm.

**Spin transport in a reference graphene lateral spin valve.** We first study the spin transport in a standard graphene LSV as shown in Fig. 2a. A spin-polarized current ($I_C$) is injected from a Co electrode into the graphene channel, creating a spin accumulation at the Co/graphene interface. This spin accumulation diffuses toward both sides of the graphene channel, creating a pure spin current, which is detected by another Co electrode as a nonlocal voltage ($V_{NL}$), see Fig. 2a. The non-local resistance $R_{NL} = V_{NL}/I_C$ is high ($R_{NL}^P$, parallel) and low ($R_{NL}^{AP}$, antiparallel) depending on the relative orientation of the magnetization of the two electrodes, which can be set by applying an in-plane magnetic field in the $x$ direction ($B_x$) due to the shape anisotropy of the electrodes (Fig. 2b). The difference $\Delta R_{NL} = R_{NL}^P - R_{NL}^{AP}$ is the spin signal. We obtain a spin signal of $\Delta R_{NL}^{ref} \sim 3 \, \Omega$ due to the large interface resistance given by good quality $TiO_2$. A Hanle measurement has been performed to characterize the spin transport properties of the graphene channel (Fig. 2c). Since the injected spins are oriented along the $x$ direction, a perpendicular in-plane magnetic field $B_y$ is applied. The precession and decoherence of the spins cause the oscillation and decay of the signal. In addition, the effect of the rotation of the Co magnetizations with $B_y$ tends to align the polarization of the injected spin current with the applied field, restoring the $R_{NL}$ signal to its zero-field value when the Co electrodes reach parallel magnetizations along the $y$ direction at high enough $B_y$. By the proper combination of the measured $R_{NL}$ curves with an initial parallel (blue circles in Fig. 2c) and antiparallel (red circles in Fig. 2c) magnetization configuration of the electrodes in the *x* direction (see Supplementary Note 1), we can obtain the rotation angle $\theta$ of the Co magnetization (Fig. 2d) and the pure spin precession and decoherence (Fig. 2e). The data in Fig. 2e can be fitted using the Hanle equation[44,45] (see Supplementary Note 1). The fitting allows us to extract the spin polarization of the Co/graphene interface $P_{ICo} = 0.068 \pm 0.001$ and the spin diffusion length of graphene $\lambda_{Gr} = 1.20 \pm 0.02$ μm. Most importantly, the reference spin signals are independent of temperature (compare the amplitude of the signals in Fig. 2b at 50 K and 2e at 300 K), in agreement with the fact that $\lambda_{Gr}$ is basically insensitive to temperature[37-39]. In contrast, the spin diffusion length of metallic channels such as Cu and Ag are significantly reduced with increasing temperature[42,46].

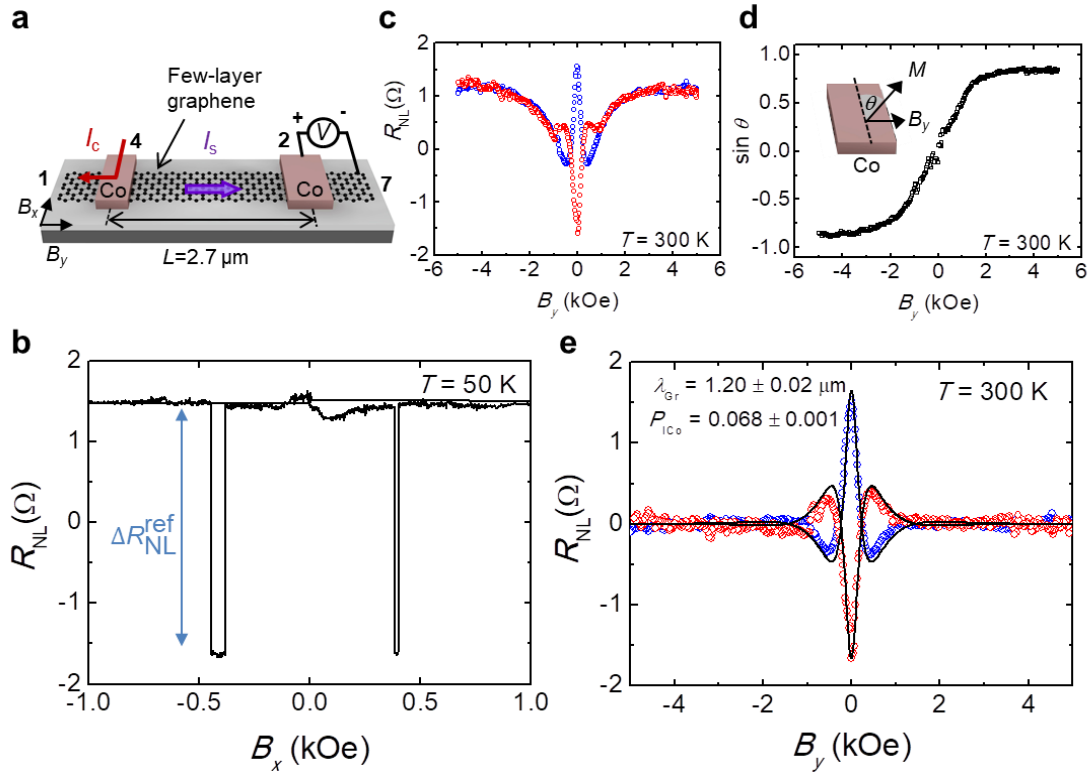

**Figure 2 | Spin transport in a reference trilayer graphene lateral spin valve.** (a) Sketch of the measurement configuration, including the electrodes from Fig. 1b that are used, and the directions of the applied magnetic field ($B_x$ and $B_y$). (b) Non-local resistance as a function of $B_x$ measured with $I_C$ = 10 μA at 50 K and center-to-center Co electrode spacing $L$ = 2.7 μm in the configuration shown in (a). $R_{NL}$ switches between high and low resistance states for parallel and antiparallel magnetization orientation of the Co electrodes while sweeping $B_x$. The reference spin signal ($\Delta R_{NL}^{ref}$) is tagged. No baseline signal has been subtracted. (c) Hanle measurement, for which $R_{NL}$ is measured in the same device as a function of $B_y$ with $I_C$ = 10 μA at 300 K in the configuration shown in (a) while the injecting and detecting Co electrodes are in the parallel (blue) and antiparallel (red) magnetization configurations. (d) sin $\theta$ as a function of $B_y$ extracted from data in (c). Inset: the magnetization direction of the Co electrode relative to *x* direction defines the angle $\theta$. (e) Pure spin precession and decoherence data extracted from data in (c), where the contribution from the in-plane magnetization rotation of the electrodes under $B_y$ is removed. Spin transport properties are extracted by fitting the Hanle equation to the experimental data (black solid lines, see Supplementary Note 1).

**Spin absorption by Pt in a graphene lateral spin valve.** Once we have extracted the spin transport properties of graphene from a reference LSV, we now explore the spin absorption by Pt in the very same device. For this experiment, we use the non-local configuration shown in Fig. 3a. A pure spin current in graphene is generated by spin injection from one Co electrode and detected by a second Co electrode, but in this case the pure spin current is partially absorbed by the Pt wire present in the middle of the spin current path before reaching the detector. The spin signal we measure after absorption by Pt is $\Delta R_{NL}^{abs} \sim 25$ mΩ, which is more than two orders of magnitude smaller than expected without the presence of the middle Pt wire (compare inset of Fig. 3b with Fig. 2b). This result indicates that the Pt wire acts as an extremely efficient spin absorber. We observe that $\Delta R_{NL}^{abs}$ has weak temperature dependence as it occurs in the reference LSV, implying that the Pt wire absorbs similar amount of spins across the temperature range investigated (see Fig. 3b).

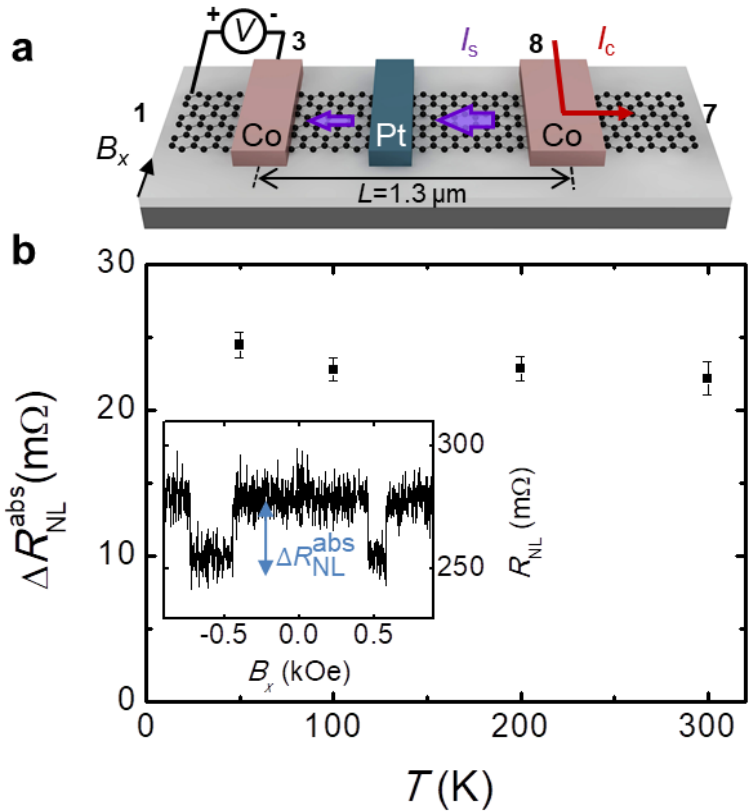

**Figure 3 | Spin absorption by Pt in a trilayer graphene lateral spin valve.** (a) Sketch of the measurement configuration, including the electrodes from Fig. 1b that are used, and the direction of the applied magnetic field ($B_x$). (b) Spin signal after Pt absorption $\Delta R_{NL}^{abs}$ as a function of temperature. Inset: Non-local resistance as a function of $B_x$ measured with $I_C$ = 10 μA and center-to-center Co electrode spacing $L$ = 1.3 μm in the configuration shown in (a), from which the values of $\Delta R_{NL}^{abs}$ are extracted for different temperatures. The curve shown corresponds to 50 K. Error bars are calculated using the standard errors associated with the statistical average of the nonlocal resistance in the parallel and antiparallel states.

**Spin generation and detection in a graphene/Pt lateral heterostructure.** After confirming that the Pt wire absorbs the spin current from graphene, and taking into account that Pt has a large $\theta_{SH}$[20,21], our next experiment demonstrates we can indeed electrically detect this spin current by using the ISHE of the same Pt wire, employing the measurement configuration in Fig. 4a (top sketch). In this case, the pure spin current injected from the Co electrode diffuses along the graphene channel and is mostly absorbed by the Pt wire. In the Pt wire, due to the ISHE, a charge current perpendicular to both the spin current direction and the spin polarization is created (Fig. 1a right) and, thus, a voltage drop is generated along the Pt wire. The measured voltage normalized to the injected current $I_C$ yields the ISHE resistance, $R_{ISHE}$. By sweeping the magnetic field ($B_y$) from positive to negative, the magnetization of the Co electrode (as well as the orientation of the spin polarization) rotates, and $R_{ISHE}$ reverses sign as shown by the blue curve in Fig. 4b. According to the symmetry of the ISHE, the signal detected in the Pt wire should be proportional to sin $\theta$ [11,13], a value which has been extracted from the Hanle data (Fig. 2d). Indeed, we observe a perfect match when overlapping the ISHE signal with sin $\theta$ as a function of $B_y$ (Fig. 4b). This excellent match unambiguously confirms that the measured signal arises from spin-to-charge conversion. Other spurious effects such as magnetoresistance or heating are ruled out with control experiments (see Supplementary Note 2). The magnitude of the spin-to-charge conversion signal $\Delta R_{SCC}$ can be calculated by taking the difference between the saturation $R_{ISHE}$ resistance at large positive and negative field ($R_{ISHE}^+ - R_{ISHE}^- = \Delta R_{SCC}$).

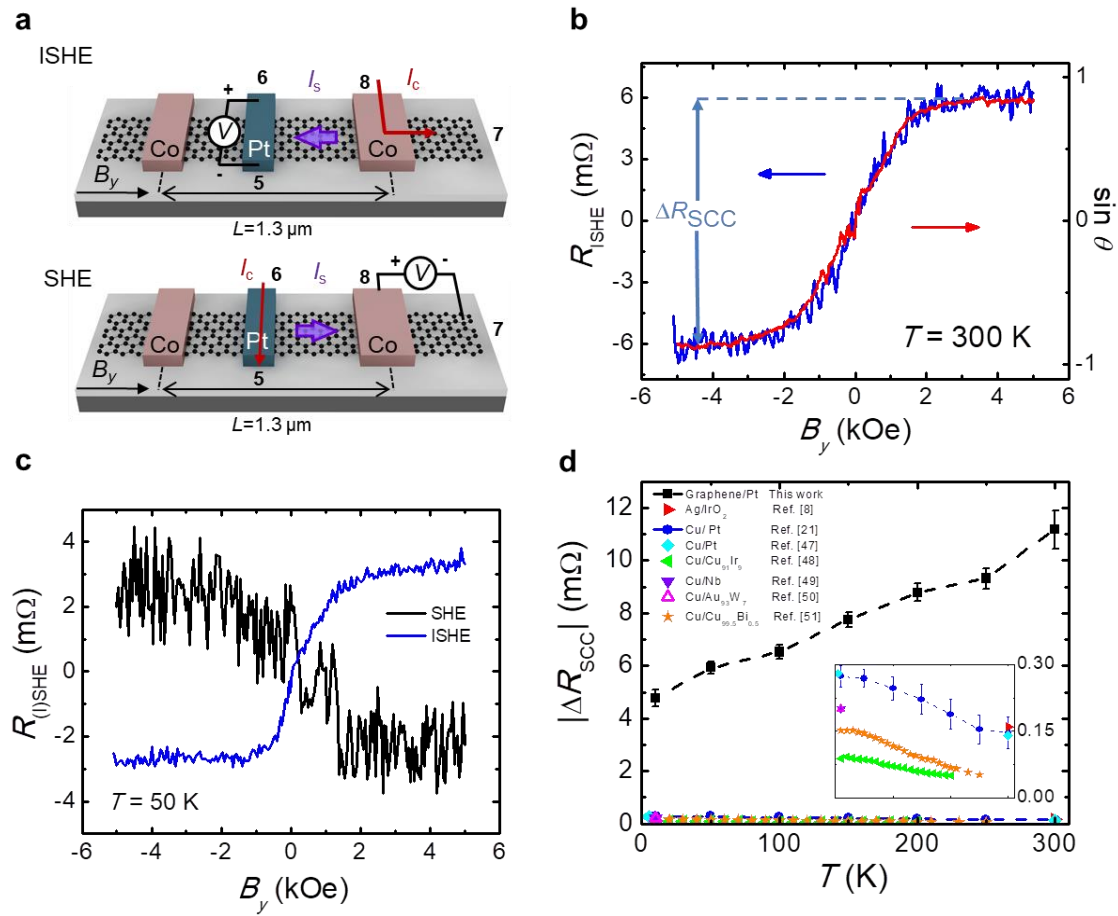

**Figure 4 | Spin-to-charge conversion in a trilayer graphene/Pt lateral heterostructure.** (a) Sketch of the ISHE (top) and the SHE (bottom) measurement configurations, including the electrodes from Fig. 1b that are used, and the direction of the applied magnetic field ($B_y$). (b) ISHE resistance (blue) as a function of $B_y$ measured with $I_C$ = 10 μA at 300 K. A baseline signal of 6.5 mΩ, corresponding to the Ohmic contribution given by the van der Pauw currents spreading into the voltage detector, has been subtracted. For comparison, sin $\theta$ (red) as a function of $B_y$ extracted from the Hanle measurement is also shown. The spin-to-charge conversion signal $\Delta R_{SCC}$ is tagged. (c) The ISHE (blue) and SHE (black) resistance as a function of $B_y$ measured with $I_C$ = 10 μA at 50 K in the configurations sketched in (a) with center-to-center Co electrode spacing $L$ = 1.3 μm, showing the reciprocity of the two effects. A baseline signal of 4 mΩ (7 mΩ), corresponding to the Ohmic contribution, has been subtracted from the ISHE (SHE) curve. (d) Experimental values of $\Delta R_{SCC}$ at different temperatures measured in the graphene/Pt heterostructure. Literature values of $\Delta R_{SCC}$ of various spin Hall metals employing different metallic spin channels are also included for comparison: Ag/IrO$_2$ (ref. 8), Cu/Pt (ref. 21 and 47), Cu/Cu$_{91}$Ir$_9$ (ref. 48), Cu/Nb (ref. 49), Cu/Au$_{93}$W$_7$ (ref. 50) and Cu/Cu$_{99.5}$Bi$_{0.5}$ (ref. 51). Inset: Zoom of the main plot showing the data of the devices with metallic spin channels. Error bars are calculated using the standard errors associated with the statistical average of the nonlocal resistance at positive and negative saturated magnetic fields.

The ISHE experiment shows that the Pt electrode can electrically detect spins travelling in the graphene channel. Next, we demonstrate that a pure spin current can also be generated using the SHE of Pt and injected into graphene. Here, we pass a charge current through the Pt wire as shown in Fig. 4a (bottom sketch). The transverse spin current generated in Pt by SHE has a spin polarization oriented along the *y* axis, and the spin accumulation in the graphene/Pt interface leads to spin injection into graphene (Fig. 1a left). By employing now the Co electrode as a detector, we are able to measure the pure spin current reaching the Co electrode as a voltage, obtaining the corresponding non-local resistance, $R_{\text{SHE}}$ (black curve in Fig. 4c). We observe that $R_{\text{SHE}}(B_y) = R_{\text{ISHE}}(-B_y)$ by swapping the voltage and current probes with the same polarity (see detailed electrode configurations in Fig. 4a), confirming the reciprocity between the ISHE and SHE in our experiment via the Onsager relation[13,52]. The SHE and ISHE measurements demonstrate that it is possible to generate and detect pure spin currents in graphene using a non-magnetic spin Hall metal.

We have performed the ISHE experiment at different temperatures, as shown in Fig. 4d. Interestingly, as the temperature is increased from 10 K to 300 K, $\Delta R_{\text{SCC}}$ increases from ~5 mΩ to ~11 mΩ, indicating that the spin-to-charge conversion signal improves at higher temperatures. This increase of $\Delta R_{\text{SCC}}$ with temperature is robust and reproducible among different samples (Supplementary Note 3). Our devices based on the few-layer graphene/Pt heterostructure show superior performance over devices reported in literature using a metallic spin channel[8,21,47-51], as summarized in Fig. 4d. Two key aspects can be highlighted. In the first place, the $\Delta R_{\text{SCC}}$ signal measured in our devices is almost two orders of magnitude larger at 300 K. In the second place, the output signal in a graphene/Pt heterostructure increases significantly with increasing temperature in contrast to the decreasing tendency found when using a metallic channel.

**Discussion**

Our experimental observations can be well explained by the standard one-dimensional spin diffusion model. The spin signal after absorption is given by the following equation (see Supplementary Note 4 for details):

$$\Delta R_{\text{NL}}^{\text{abs}} = 8 R_{\text{Gr}} Q_{\text{ICo1}} Q_{\text{ICo2}} P_{\text{ICo}}^2 \frac{(Q_{\text{IPt}}+Q_{\text{Pt}})e^{-\frac{L}{\lambda_{\text{Gr}}}}}{(2Q_{\text{ICo1}}+1)(2Q_{\text{ICo2}}+1)-2(Q_{\text{ICo1}}+Q_{\text{ICo2}}+1)e^{-\frac{L}{\lambda_{\text{Gr}}}}+e^{-\frac{2L}{\lambda_{\text{Gr}}}}}, \qquad (1)$$

where $Q_{Ik} = \frac{1}{1-P_{Ik}^2}\frac{R_{Ik}}{R_{\text{Gr}}}$, being $R_{Ik}$ the resistance and $P_{Ik}$ the spin polarization of the $k^{\text{th}}$ metal/graphene interface. In our device, we define $k$ = Co1, Co2, Pt for the Co injector, Co detector and Pt wire, respectively, and we assume $P_{\text{ICo1}} = P_{\text{ICo2}} = P_{\text{ICo}}$. $R_{\text{Gr}} = \frac{R_{\text{Gr}}^{\blacksquare}\lambda_{\text{Gr}}}{w_{\text{Gr}}}$ is the spin resistance of graphene, where $R_{\text{Gr}}^{\blacksquare}$ is its sheet resistance. $Q_{\text{Pt}} = \frac{R_{\text{Pt}}}{R_{\text{Gr}}}$, being $R_{\text{Pt}} = \frac{\rho_{\text{Pt}}\lambda_{\text{Pt}}}{w_{\text{Pt}}w_{\text{Gr}}\tanh[t_{\text{Pt}}/\lambda_{\text{Pt}}]}$ the spin resistance of Pt. The geometrical factors $w_{\text{Gr}}$, $w_{\text{Pt}}$, $t_{\text{Pt}}$ and $L$ are the width of graphene, width of Pt, thickness of Pt and center-to-center distance between the Co electrodes, respectively. $\lambda_{\text{Pt}}$ is the spin diffusion length of Pt.

The spin-to-charge conversion signal $\Delta R_{\text{SCC}}$ of the ISHE experiment is given by the following expression[21,48,49]:

$$\Delta R_{\text{SCC}} = \frac{2\theta_{\text{SH}}\rho_{\text{Pt}}x_{\text{Pt/Gr}}}{w_{\text{Pt}}}\left(\frac{\bar{I}_S}{I_C}\right), \qquad (2)$$

where $\bar{I}_S$ is the effective spin current injected vertically from graphene into the Pt wire that contributes to the ISHE in Pt and $x_{Pt/Gr}$ is the correction factor which considers the current in the Pt shunted through the graphene (see Supplementary Note 5 for details).

For the calculation, we substitute into equation (1) and (2) the experimental values of $\Delta R_{NL}^{abs}$ and $\Delta R_{SCC}$, the obtained $P_{ICo}$ and $\lambda_{Gr}$ (Supplementary Note 1), the geometrical factors (measured from SEM images), and the values of $\rho_{Pt}$ and $\theta_{SH}$ of Pt[21]. We assume negligible current shunting into the graphene due to the much larger sheet resistance of graphene when compared to Pt at the junction area, $R_{Gr}^{\blacksquare} = 1085\ \Omega$ vs $\rho_{Pt}/t_{Pt} = 64\ \Omega$ at 300 K, which leads to $x_{Pt/Gr} \approx 1$. We extract two very sensitive parameters $\lambda_{Pt}$ and $R_{IPt}$, which are 2.1 ± 0.4 nm and 8.4 ± 0.4 Ω at 300 K. The obtained $\lambda_{Pt}$ is expected when considering the resistivity of our Pt wire[21]. The small value of $R_{IPt}$ facilitates strong spin absorption by Pt from graphene and is compatible with our direct measurement (Supplementary Note 4). The good consistency of extracted values confirms that our assumption of $x_{Pt/Gr} \approx 1$ is robust.

Having quantified accurately all the parameters in our system, we can confirm the origin of the observed large spin-to-charge conversion and its strong temperature dependence. It mainly arises from four factors. First, the superior spin transport properties of graphene ($\lambda_{Gr}$ ~1.2 µm) and its temperature insensitivity. Graphene's exceptional ability to transport spins remains intact at room temperature, i.e. the same amount of spin current arrives to the Pt absorber at different temperatures; Second, while the amount of spin current to be converted remains the same, the efficiency of the conversion ($\theta_{SH}$) of Pt increases linearly with temperature from 17.8 ± 2.0% at 50 K to 23.4 ± 2.5% at 300 K (ref. 21); Third, the resistivity of Pt increases from 99 µΩcm at 50 K to 134 µΩcm at 300 K; and fourth, the negligible shunting of the charge current in Pt by graphene ($x_{Pt/Gr} \approx 1$). The enhancement of $\Delta R_{SCC}$ with increasing temperature mainly benefits from the first three factors, which are constant $\lambda_{Gr}$ and increasing $\theta_{SH}\rho_{Pt}$ product as described in equation (2). In contrast, in metallic spin channels, the spin diffusion length of the metal channel decreases significantly with increasing temperature[42,46], leading to reduced output voltage. Our devices give much larger $\Delta R_{SCC}$ than those using metallic spin channels mainly due to the first (long spin diffusion length of graphene) and fourth (negligible shunting) factors. In traditional metallic spin valve devices, the resistivity of the metal channel is close or smaller than that of the spin Hall metal, thus $x$ are much lower (0.05-0.36)[21,51], a serious issue preventing large spin-to-charge conversion pointed out recently[8]. However, in our device with few-layer graphene/Pt heterostructure, $x_{Pt/Gr} \approx 1$ is close to ideal and the use of more resistive graphene (single or bilayer) is not necessary, since $x_{Pt/Gr}$ cannot be further increased. Further improvement to the spin-to-charge conversion could be easily achieved by using high quality graphene devices, where almost two orders of magnitude enhancement of $\lambda_{Gr}$ is obtained[31,32], or reducing the spin current dilution into the Pt wire by decreasing its thickness (as can be deduced from Supplementary Equation 10).

Finally, it is worth mentioning that a direct comparison between $\Delta R_{NL}^{ref}$ (~3 Ω) and $\Delta R_{SCC}$ (~12 mΩ) is not appropriate, because they quantify different outputs. Whereas the former only probes the spin accumulation in the channel through a ferromagnetic tunnel barrier and acts as spin detector, the latter is a measurement of the converted charge current through a transparent interface, which can be potentially utilized (See Supplementary Note 6 for an extended discussion).

To conclude, we succeeded in electrically injecting and detecting pure spin currents in few-layer graphene by employing the SHE and ISHE of a non-magnetic material, respectively. The extraordinary ability of graphene to transport spins, together with its relatively high resistance compared to Pt, results in the largest spin-to-charge conversion signal reported so far. Most importantly, the largest conversion, which is two orders of magnitude larger than in devices employing metallic spin channels, occurs at room temperature. The fuse and perfect match of these two elements in a heterostructural device of graphene/Pt provides new plausible opportunities for future spin-orbit-based devices.

**Methods**

**Device fabrication**. To fabricate our devices, few-layer graphene flakes are first produced by micromechanical cleavage of natural graphite onto 300-nm-thick $SiO_2$ on doped Si substrate using Nitto tape (Nitto SPV 224P) and identified using its optical contrast[41]. We select flakes with the most convenient shape (long and narrow), regardless of the number of layers, since the excellent spin transport properties do not depend strongly on the number of graphene layers[36]. The nanofabrication of the device follows two steps of e-beam lithography with electrode metal deposition and lift-off. For the 200-nm-wide Pt wires, 21 nm of Pt were sputtered at 0.6 Ås$^{-1}$ using 40 W in 3 mTorr of Ar pressure. This deposition condition gives rise to very resistive Pt with $\rho_{Pt}$ = 99 (134) μΩcm at 50 (300) K. The 35-nm-thick Co electrodes with widths between 150 and 350 nm are deposited in an ultra-high vacuum chamber using e-beam evaporation on top of 6 Å of Ti after the natural oxidation of Ti in air. The presence of $TiO_2$ between the Co electrode and the graphene channel leads to interface resistances between 10 and 42 kΩ.

**Electrical measurements.** The measurements are performed in a Physical Property Measurement System (PPMS) by Quantum Design, using a DC reversal technique with a Keithley 2182 nanovoltmeter and a 6221 current source[42-44].

**Data availability**. All relevant data are available from the authors.

**References**

1. Datta, S. & Das, B. Electronic analog of the electro-optic modulator. *Appl. Phys. Lett*. **56**, 665-667 (1990).
2. Wunderlich, J. *et al.* Spin Hall effect transistor. *Science* **330**, 1801-1804 (2010).
3. Choi, W. Y. *et al.* Electrical detection of coherent spin precession using the ballistic intrinsic spin Hall effect. *Nat. Nanotechnol*. **10**, 666-670 (2015).
4. Yan, W. *et al.* A two-dimensional spin field-effect switch. *Nat. Commun.* **7**, 13372 (2016).
5. Liu, L. *et al.* Spin-torque switching with the giant spin Hall effect of tantalum. *Science* **336**, 555-558 (2012).
6. Miron, I. M. *et al.* Perpendicular switching of a single ferromagnetic layer induced by in-plane current injection. *Nature* **476**, 189-193 (2011).
7. Garello, K. *et al.* Symmetry and magnitude of spin-orbit torques in ferromagnetic heterostructures. *Nat. Nanotechnol*. **8**, 587-593 (2013).
8. Fujiwara, K. *et al.* 5d iridium oxide as a material for spin-current detection. *Nat. Commun.* **4**, 2893 (2013).
9. Kuschel, T. & Reiss, G. Spin orbitronics: Charges ride the spin wave. *Nat. Nanotechnol.* **10**, 22-24 (2015).
10. Manchon, A., Koo, H. C., Nitta, J., Frolov, S. M. & Duine, R. A. New perspectives for Rashba spin-orbit coupling. *Nat. Mater*. **14**, 871-882 (2015).
11. Valenzuela, S. O. & Tinkham, M. Direct electronic measurement of the spin Hall effect. *Nature* **442**, 176-179 (2006).
12. Saitoh, E., Ueda, M., Miyajima H. & Tatara, G. Conversion of spin current into charge current at room temperature: Inverse spin-Hall effect. *Appl. Phys. Lett*. **88**, 182509 (2006).
13. Kimura, T., Otani, Y., Sato, T., Takahashi, S. & Maekawa, S. Room-temperature reversible spin Hall effect. *Phys. Rev. Lett*. **98**, 156601 (2007).
14. Rojas Sánchez, J. C. *et al.* Spin-to-charge conversion using Rashba coupling at the interface between non-magnetic materials. *Nat. Commun.* **4**, 2944 (2013).
15. Isasa, M. *et al.* Origin of inverse Rashba-Edelstein effect detected at the Cu/Bi interface using lateral spin valves. *Phys. Rev. B* **93**, 014420 (2016).

16. Vaklinova, K., Hoyer, A., Burghard, M. & Kern, K. Current-induced spin polarization in topological insulator-graphene heterostructures. *Nano Lett*. **16**, 2595-2602 (2016).
17. Dankert, A., Geurs, J., Kamalakar, M. V., Charpentier, S. & Dash, S. P. Room temperature electrical detection of spin polarized currents in topological insulators. *Nano Lett.* **15**, 7976-7981 (2015).
18. Manipatruni, S., Nikonov, D. E. & Young, I. A. Spin-orbit logic with magnetoelectric nodes: a scalable charge mediated nonvolatile spintronic logic. Preprint at https://arxiv.org/abs/1512.05428 (2017).
19. Sinova, J., Valenzuela, S. O., Wunderlich, J., Back, C. H. & Jungwirth, T. Spin Hall effects. *Rev. Mod. Phys.* **87**, 1213-1259 (2015).
20. Nguyen, M. H., Ralph, D. C. & Buhrman, R. A. Spin torque study of the spin Hall conductivity and spin diffusion length in platinum thin films with varying resistivity. *Phys. Rev. Lett*. **116**, 126601 (2016).
21. Sagasta, E. *et al*. Tuning the spin Hall effect of Pt from the moderately dirty to the superclean regime. *Phys. Rev. B* **94**, 060412(R) (2016).
22. Pai, C.-F. *et al*. Spin transfer torque devices utilizing the giant spin Hall effect of tungsten. *Appl. Phys. Lett.* **101**, 122404 (2012).
23. Demasius, K.-U *et al*. Enhanced spin-orbit torques by oxygen incorporation in tungsten films. *Nat. Commun*. **7**, 10644 (2016).
24. Pesin, D. & MacDonald, A. H. Spintronics and pseudospintronics in graphene and topological insulators. *Nat. Mater*. **11**, 409-416 (2012).
25. Friedman, A. L., van't Erve, O. M. J., Robinson, J. T., Whitener, Jr., K. E. & Jonker, B. T. Hydrogenated graphene as a homoepitaxial tunnel barrier for spin and charge transport in graphene. *ACS Nano* **9**, 6747-6755 (2015).
26. Lin, C.-C. *et al*. Improvement of spin transfer torque in asymmetric graphene devices. *ACS Nano* **8**, 3807-3812 (2014).
27. Han, W. *et al*. Tunneling spin injection into single layer graphene. *Phys. Rev. Lett*. **105**, 167202 (2010).
28. Raes, B. *et al*. Determination of the spin-lifetime anisotropy in graphene using oblique spin precession. *Nat. Commun.* **7**, 11444 (2016).
29. Kamalakar, M. V., Groenveld, C., Dankert, A. & Dash, S. P. Long distance spin communication in chemical vapour deposited graphene. *Nat. Commun*. **6**, 6766 (2015).
30. Avsar, A. *et al*. Toward Wafer Scale Fabrication of Graphene Based Spin Valve Devices. *Nano Lett*. **11**, 2363-2368 (2011).
31. Drögeler, M. *et al*. Spin Lifetimes Exceeding 12 ns in Graphene Nonlocal Spin Valve Devices. *Nano Lett*. **16**, 3533-3539 (2016).
32. Ingla-Aynés, J., Meijerink, R. J. & van Wees B. J. Eighty-Eight Percent Directional Guiding of Spin Currents with 90 μm Relaxation Length in Bilayer Graphene Using Carrier Drift. *Nano Lett.* **16**, 4825-4830 (2016).
33. Wen, H. *et al*. Experimental Demonstration of XOR Operation in Graphene Magnetologic Gates at Room Temperature. *Phys. Rev. Appl*. **5**, 044003 (2016).
34. Yan, W. *et al*. Long Spin Diffusion Length in Few-Layer Graphene Flakes. *Phys. Rev. Lett*. **117**, 147201 (2016).
35. Drögeler, M. *et al*. Nanosecond Spin Lifetimes in Single- and Few-Layer Graphene–hBN Heterostructures at Room Temperature. *Nano Lett*. **14**, 6050-6055 (2014).
36. Maassen, T., Dejene, F. K., Guimarães, M. H. D., Józsa, C. & van Wees, B. J. Comparison between charge and spin transport in few-layer graphene. *Phys. Rev. B* **83**, 115410 (2011).
37. Tombros, N., Jozsa, C., Popinciuc, M., Jonkman, H. T. & van Wees, B. J. Electronic spin transport and spin precession in single graphene layers at room temperature. *Nature* **448**, 571-574 (2007).
38. Yang, T.-Y. *et al*. Observation of Long Spin-Relaxation Times in Bilayer Graphene at Room Temperature. *Phys. Rev. Lett*. **107**, 047206 (2011).
39. Ertler, C., Konschuh, S., Gmitra, M. & Fabian, J. Electron spin relaxation in graphene: The role of the substrate. *Phys. Rev. B* **80**, 041405(R) (2009).
40. Cubukcu, M. *et al*. Ferromagnetic tunnel contacts to graphene: Contact resistance and spin signal. *J. Appl. Phys.* **117**, 083909 (2015).
41. Castellanos-Gomez, A. *et al*. Deterministic transfer of two-dimensional materials by all-dry viscoelastic stamping. *2D Mater.* **1**, 011002 (2014).
42. Villamor, E., Isasa, M., Hueso, L. E. & Casanova, F. Contribution of defects to the spin relaxation in copper nanowires. *Phys. Rev. B* **87**, 094417 (2013).
43. Erekhinsky, M., Casanova, F., Schuller, I. K. & Sharoni, A. Spin-dependent Seebeck effect in non-local spin valve devices. *Appl. Phys. Lett.* **100**, 212401 (2012).
44. Villamor, E., Hueso, L. E. & Casanova, F. Effect of the interface resistance in non-local Hanle measurements. *J. Appl. Phys.* **117**, 223911 (2015).

45. Idzuchi, H., Fert, A. & Otani, Y. Revisiting the measurement of the spin relaxation time in graphene-based devices. *Phys. Rev. B* **91**, 241407(R) (2015).
46. Fukuma, Y. *et al*. Giant enhancement of spin accumulation and long-distance spin precession in metallic lateral spin valves. *Nat. Mater*. **10**, 527-531 (2011).
47. Vila, L., Kimura, T. & Otani, Y. Evolution of the Spin Hall Effect in Pt Nanowires: Size and Temperature Effects. *Phys. Rev. Lett*. **99**, 226604 (2007).
48. Niimi, Y. *et al*. Extrinsic Spin Hall Effect Induced by Iridium Impurities in Copper. *Phys. Rev. Lett*. **106**, 126601 (2011).
49. Morota, M. *et al*. Indication of intrinsic spin Hall effect in 4d and 5d transition metals. *Phys. Rev. B* **83**, 174405 (2011).
50. Laczkowski, P. *et al*. Experimental evidences of a large extrinsic spin Hall effect in AuW alloy. *Appl. Phys. Lett*. **104**, 142403 (2014).
51. Niimi, Y. *et al*. Giant Spin Hall Effect Induced by Skew Scattering from Bismuth Impurities inside Thin Film CuBi Alloys. *Phys. Rev. Lett*. **109**, 156602 (2012).
52. Due to the breaking of the time-reveral symmetry when a magnetic field is applied, the Onsager's reciprocity relation reads $L_{ij}(B) = L_{ji}(-B)$ where the linear coefficient $L_{ij}$ determines the response of the flux density $J_i$ (for instance the spin current) to a weak thermodynamic force $X_j$ (for instance the electric field)[53]. For SHE and ISHE, $L_{ij}$ is the non-local resistance and, therefore, $R_{\text{SHE}}(B_y) = R_{\text{ISHE}}(-B_y)$.
53. Jacquod, P., Whitney, R. S., Meair, J. & Büttiker M. Onsager relations in coupled electric, thermoelectric, and spin transport: The tenfold way. *Phys. Rev. B* **86**, 155118 (2012).

## Acknowledgements

This work was supported by the European Union 7th Framework Programme under the Marie Curie Actions (607904-13-SPINOGRAPH), by the Spanish MINECO under Project No. MAT2015-65159-R, and by the Basque Government under Project No. PC2015-1-01. E.S. thanks the Spanish MECD for a Ph.D. fellowship (Grant No. FPU14/03102).

## Author Contributions

F.C. conceived the study. W.Y., E.S. and M.R. performed the experiments. W.Y., E.S., L.E.H. and F.C. analysed the data, discussed the experiments and wrote the manuscript; Y.N. derived the equations used in the analysis of the spin absorption technique. All the authors contributed to the scientific discussion and manuscript revision. L.E.H. and F.C. co-supervised the work.

# SUPPLEMENTARY INFORMATION

**Supplementary Note 1: Spin transport properties of graphene**

The spin transport properties of graphene are obtained from the fitting of the Hanle measurement. The decoherence of the spin during precession causes the decay of an oscillating signal, which can be fitted using the following equation[1]:

$$R_{NL} = -2R_N \left(\frac{P_{F1}}{1-P_{F1}^2}\frac{R_{F1}}{R_N} + \frac{P_{I1}}{1-P_{I1}^2}\frac{R_{I1}}{R_N}\right)\left(\frac{P_{F2}}{1-P_{F2}^2}\frac{R_{F2}}{R_N} + \frac{P_{I2}}{1-P_{I2}^2}\frac{R_{I2}}{R_N}\right)\frac{C_{12}}{\det(\check{X})}, \qquad (1)$$

where $R_{Fk} = \rho_F \lambda_F / A_{Ik}$ are the spin resistances of the $k^{th}$ FM contact ($k = 1$ is the injector and $k = 2$ is the detector), with resistivity $\rho_F$, spin diffusion length $\lambda_F$ and contact area $A_{Ik}$; $P_{Fk}$ are the spin polarizations of the $k^{th}$ FM contact; $R_N = \frac{R_{Gr}^\blacksquare \lambda_{Gr}}{w_{Gr}}$ is the spin resistance of graphene calculated with its sheet resistance ($R_{Gr}^\blacksquare$), spin diffusion length ($\lambda_{Gr}$) and width ($w_{Gr}$); $R_{Ik} = 1/G_{Ik}$ is the resistance of the $k^{th}$ interface, where $G_{Ik} = G_{Ik}^\uparrow + G_{Ik}^\downarrow$ is the conductance of the $k^{th}$ interface that considers both spin up and down channels; $P_{Ik} = (G_{Ik}^\uparrow - G_{Ik}^\downarrow)/(G_{Ik}^\uparrow + G_{Ik}^\downarrow)$ describes the interfacial spin polarization; and $C_{12}$ and $\det(\check{X})$ are defined as[1]:

$$C_{12} = -\det\begin{pmatrix} \mathrm{Re}[\bar{\lambda}_\omega e^{-L/\tilde{\lambda}_\omega}] & -\mathrm{Im}[\bar{\lambda}_\omega e^{-L/\tilde{\lambda}_\omega}] & -\mathrm{Im}[\bar{\lambda}_\omega] \\ \mathrm{Im}[\bar{\lambda}_\omega] & r_{1\perp} + \mathrm{Re}[\bar{\lambda}_\omega] & \mathrm{Re}[\bar{\lambda}_\omega e^{-L/\tilde{\lambda}_\omega}] \\ \mathrm{Im}[\bar{\lambda}_\omega e^{-L/\tilde{\lambda}_\omega}] & \mathrm{Re}[\bar{\lambda}_\omega e^{-L/\tilde{\lambda}_\omega}] & r_{2\perp} + \mathrm{Re}[\bar{\lambda}_\omega] \end{pmatrix}, \qquad (2)$$

$$\check{X} = \begin{pmatrix} r_{1\|} + \mathrm{Re}[\bar{\lambda}_\omega] & \mathrm{Re}[\bar{\lambda}_\omega e^{-L/\tilde{\lambda}_\omega}] & -\mathrm{Im}[\bar{\lambda}_\omega] & -\mathrm{Im}[\bar{\lambda}_\omega e^{-L/\tilde{\lambda}_\omega}] \\ \mathrm{Re}[\bar{\lambda}_\omega e^{-L/\tilde{\lambda}_\omega}] & r_{2\|} + \mathrm{Re}[\bar{\lambda}_\omega] & -\mathrm{Im}[\bar{\lambda}_\omega e^{-L/\tilde{\lambda}_\omega}] & -\mathrm{Im}[\bar{\lambda}_\omega] \\ \mathrm{Im}[\bar{\lambda}_\omega] & \mathrm{Im}[\bar{\lambda}_\omega e^{-L/\tilde{\lambda}_\omega}] & r_{1\perp} + \mathrm{Re}[\bar{\lambda}_\omega] & \mathrm{Re}[\bar{\lambda}_\omega e^{-L/\tilde{\lambda}_\omega}] \\ \mathrm{Im}[\bar{\lambda}_\omega e^{-L/\tilde{\lambda}_\omega}] & \mathrm{Im}[\bar{\lambda}_\omega] & \mathrm{Re}[\bar{\lambda}_\omega e^{-L/\tilde{\lambda}_\omega}] & r_{2\perp} + \mathrm{Re}[\bar{\lambda}_\omega] \end{pmatrix}, \qquad (3)$$

where $\bar{\lambda}_\omega = \tilde{\lambda}_\omega/\lambda_N$ with $\tilde{\lambda}_\omega = \frac{\lambda_N}{\sqrt{i+i\omega_L\tau_{sf}}}$ and the Larmor frequency $\omega_L = \gamma_e B_\perp = \frac{g\mu_B}{\hbar}B_\perp$; $r_{k\|} = \left(\frac{2}{1-P_{Ik}^2}\frac{R_{Ik}}{R_N} + \frac{2}{1-P_{Fk}^2}\frac{R_{Fk}}{R_N}\right)$; $L$ is the center-to-center distance between FM electrodes; and $r_{k\perp} = \frac{1}{R_N G_{Ik}^{\uparrow\downarrow}}$ with $G_{Ik}^{\uparrow\downarrow}$ being the spin mixing interface conductance.

The Hanle measurement of the reference graphene LSV (Fig. 2c of the main text) also contains the effect of the rotation of the Co magnetizations with the external magnetic field ($B_y$), which tends to align the spin polarization with $B_y$, restoring the $R_{NL}$ signal to its zero-field value $R_{NL}(0)$ for parallel Co magnetizations. When this effect is taken into account, $R_{NL}$ can be expressed as[2,3]:

$$R_{NL}^{P(AP)}(B_y, \theta) = \pm R_{NL}^P(B_y)\cos^2(\theta) + |R_{NL}(0)|\sin^2(\theta), \qquad (4)$$

where $R_{\text{NL}}^{\text{P(AP)}}$ is the non-local resistance measured as a function of $B_y$ when the two Co electrodes are parallel (P) or antiparallel (AP) and $\theta$ is the angle of the Co magnetization with respect to the easy axis of the electrode ($x$ axis). Note that the sign "+" corresponds to the P curve and "–" to the AP curve, and that $R_{\text{NL}}^{\text{P}}(B_y) = -R_{\text{NL}}^{\text{AP}}(B_y)$ for the pure spin precession and decoherence. By the proper combination of the measured P and AP curves, we can obtain the rotation of the Co magnetization (Fig. 2d of the main text):

$$\sin^2(\theta) = \frac{R_{\text{NL}}^{\text{P}}(B_y,\theta)+R_{\text{NL}}^{\text{AP}}(B_y,\theta)}{2|R_{\text{NL}}(0)|}, \tag{5}$$

and the pure spin precession and decoherence (Fig.2e of the main text):

$$R_{\text{NL}}^{\text{P}}(B_y) = |R_{\text{NL}}(0)|\frac{R_{\text{NL}}^{\text{P}}(B_y,\theta)-R_{\text{NL}}^{\text{AP}}(B_y,\theta)}{2|R_{\text{NL}}(0)|-R_{\text{NL}}^{\text{P}}(B_y,\theta)-R_{\text{NL}}^{\text{AP}}(B_y,\theta)}. \tag{6}$$

For the fitting of the pure spin precession and decoherence curve of the reference graphene LSV in Fig. 2e, we assume the injecting and detecting electrodes have the same spin polarizations ($P_{\text{Co1}} = P_{\text{Co2}} = P_{\text{Co}}$ and $P_{\text{ICo1}} = P_{\text{ICo2}} = P_{\text{ICo}}$) and, following ref. 1, we assume an isotropic spin absorption, hence $G_{Ik}^{\uparrow\downarrow} = 1/(2R_{Ik} + 2R_{Fk})$. We fix the following experimental parameters: $P_{\text{Co}} = 0.12$ (ref. 4), $R_{\text{ICo1}} = 42$ k$\Omega$, $R_{\text{ICo2}} = 10$ k$\Omega$, $L = 2.7$ μm, $w_{\text{Gr}} = 250$ nm, $w_{\text{Co1}} = 344$ nm, $w_{\text{Co2}} = 315$ nm, $R_{\text{Gr}}^{\blacksquare} = 1085$ $\Omega$, $\rho_{\text{Co}} = 19$ μ$\Omega$cm (ref. 4), $\lambda_{\text{Co}} = 40$ nm (ref. 5,6), and obtain $P_{\text{ICo}} = 0.068 \pm 0.001$, $D = 0.005$ m$^2$s$^{-1}$, and $\lambda_{\text{Gr}} = 1.20 \pm 0.02$ μm. Because the spin signal is constant across the temperature range from 10 K to 300 K, we assume the spin diffusion length of graphene is independent of temperature.

**Supplementary note 2: Control experiments**

In order to rule out any spurious magnetoresistance effect in graphene as the origin of the observed ISHE signal, we fabricated a control device where we substitute the Pt wire with a Cu wire, which has a weak spin-orbit coupling and, therefore, no spin-to-charge conversion signal is expected[7,8]. As the dimensions of the control device are very similar to those of the Pt/graphene device in the main manuscript, any spurious effect other than the ISHE signal, such as magnetoresistive effects arising from the stray fields created by the Co injector, should also be present in the control measurement. First of all, we check that the Co electrode is of similar quality as the Pt/graphene sample by measuring a reference spin valve in a nonlocal configuration. The nice and clear nonlocal spin signal indicates that the Co electrode next to the Cu wire is an efficient spin injector. Next, we measure the voltage drop across the Cu wire while using the Co electrode for spin injection in the ISHE measurement configuration. The result is shown in Supplementary Fig. 1b. This measurement produces a flat nonlocal background much smaller than that of the ISHE signal measured in the device presented in the main text (Device #1), indicating there is no spurious contribution to the ISHE signal (compare black and blue curves in Supplementary Fig. 1b). The same control experiments were carried out in a total of five control devices and all of them showed a similar flat background.

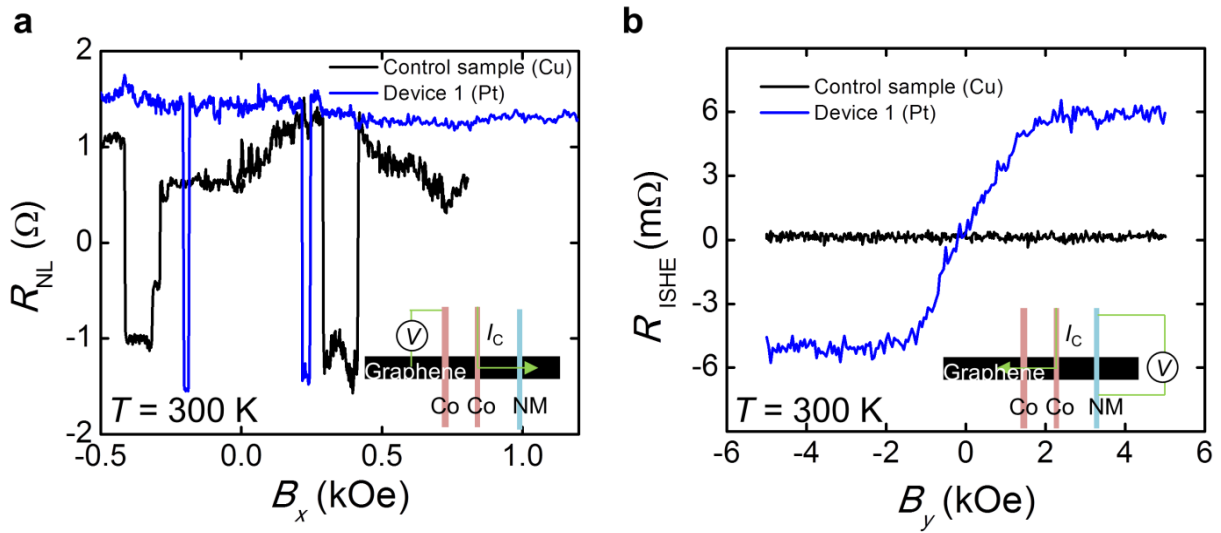

**Supplementary Figure 1 | Comparison between Device #1 and a control device using graphene lateral spin valves and Cu wire.** (a) Nonlocal magnetoresistance as a function of $B_x$ using a reference spin valve next to the non-magnetic metal wire shows similar spin signal in both the control sample and Device #1, indicating that Co electrode next to Cu wire in the control device is a spin injector as good as in Device #1. $I_C$ = 10 µA and $T$ = 300 K in both measurements. (b) ISHE resistance as a function of $B_y$ measured in both the control device and Device #1. $I_C$ = 10 µA and $T$ = 300 K in both measurements.

In order to rule out further spurious effects such as drift due to heating, we performed trace and retrace of the ISHE measurements by increasing and decreasing the applied magnetic field for each temperature. An example of this measurement from Device #1 at 300 K is shown in Supplementary Fig. 2. The overlapping of the trace and retrace curves rules out any drift due to heating.

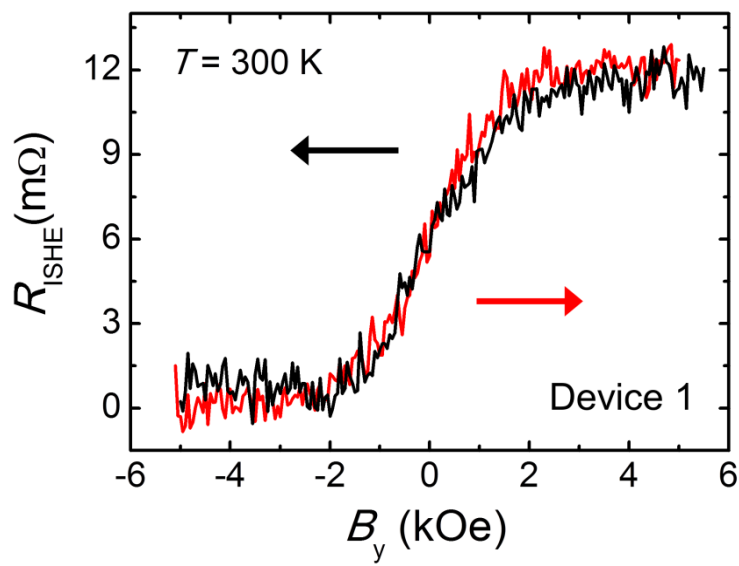

**Supplementary Figure 2 | ISHE resistance for increasing and decreasing magnetic field.** ISHE resistance as a function of $B_y$ with $I_C$ = 10 µA and $T$ = 300 K measured in Device #1. Red solid curve corresponds to the increasing field sweep and black solid curve to the decreasing field sweep. Note that the figure plots the raw data without any baseline subtraction.

**Supplementary Note 3: Reproducibility**

The key results we presented in this manuscript, which is the SHE and ISHE effect in Pt and its temperature dependence, are fully reproducible among different samples. Here, we show a second device (#2), in which we measure the SHE and ISHE effects at 300 K by simply swapping the current and voltage probes, demonstrating that they are reciprocal to each other (see Supplementary Fig. 3a). The $\Delta R_{SCC}$ signal increases with temperature, reaching the maximum value at 300 K (Supplementary Fig. 3b).

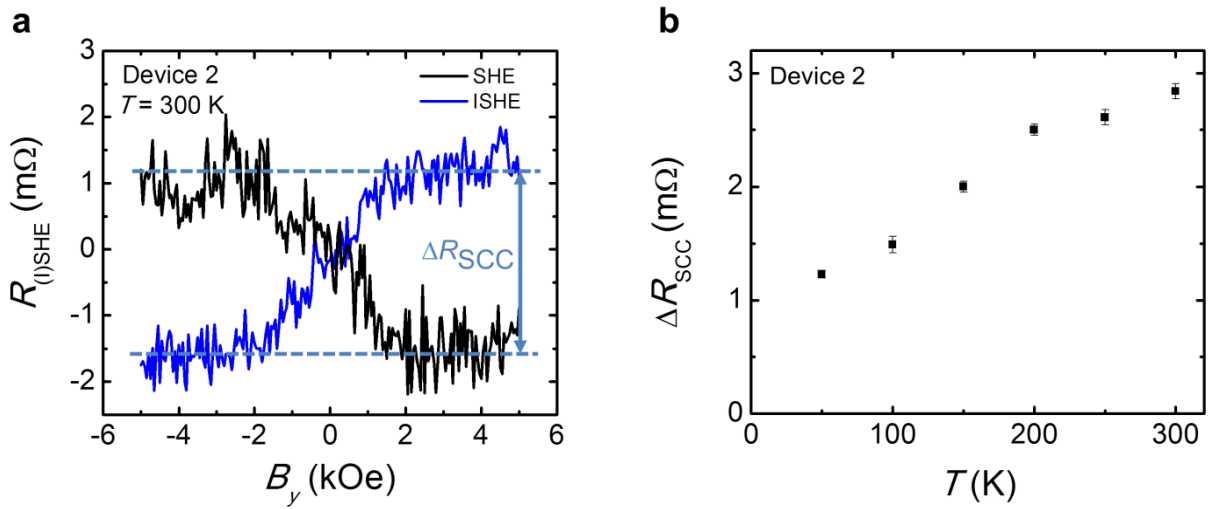

**Supplementary Figure 3 | Spin-to-charge conversion signal in another graphene/Pt lateral heterostructure.** (a) SHE (black) and ISHE (blue) resistance as a function of $B_y$ measured with $I_C$ = 10 µA at 300 K in Device #2, demonstrating the reciprocity of these two effects. A baseline signal of 19 mΩ (18 mΩ) has been subtracted from the ISHE (SHE) curve. (b) Temperature dependence of the $\Delta R_{SCC}$ signal in Device #2. Error bars are calculated using the standard errors associated with the statistical average of the nonlocal resistance at positive and negative saturated magnetic fields.

The magnitude of the $\Delta R_{SCC}$ signal measured in Device #2 is smaller than the device presented in the main text (#1) (compare Supplementary Fig. 3b with Fig. 4d), although the dimensions are very similar (the widths of the Co injector and the Pt wire in Device #2 are 336 nm and 193 nm, respectively). This is due to the variation of the interface resistance between Co and graphene, i.e., $R_{ICo1}$ in Device #2 is smaller (2 kΩ) than in Device #1 (14.7 kΩ). Despite this variation, both the $\Delta R_{SCC}$ signal and its temperature dependence can be fully explained with the spin diffusion model (Supplementary Equations 8 and 9), evidencing the robustness of the performance of the graphene/Pt lateral heterostructures.

**Supplementary Note 4: Spin absorption by Pt**

The Pt wire placed in between the two Co electrodes has a much smaller spin resistance than the graphene channel. Therefore, the presence of Pt will cause an additional relaxation to the spins transported in the graphene channel. This additional spin relaxation, which shows up as a decrease of measured spin signal, is called spin absorption. According to the standard one-dimensional (1D) spin diffusion model[9], the measured spin signal after spin absorption can be calculated using the following equation:

$$\Delta R_{NL}^{abs} = 4R_N \hat{\alpha}_1 \hat{\alpha}_2 \frac{(r_3-1)e^{-\frac{L}{\lambda_N}}}{r_1 r_2 (r_3 - Q_{I3}) - r_1(1+Q_{I3})e^{-\frac{2(L-d)}{\lambda_N}} - r_2(1-Q_{I3})e^{-\frac{2d}{\lambda_N}} - (r_3 - Q_{I3})e^{-\frac{2L}{\lambda_N}} + 2e^{-\frac{2L}{\lambda_N}}}, \quad (7)$$

with $r_k = 2Q_{Ik} + 2Q_{Fk} + 1$, $\hat{\alpha}_k = P_{Ik} Q_{Ik} + P_{Fk} Q_{Fk}$, $Q_{Ik} = \frac{1}{1-P_{Ik}^2} \frac{R_{Ik}}{R_N}$, and $Q_{Fk} = \frac{1}{1-P_{Fk}^2} \frac{R_{Fk}}{R_N}$, where $k = 1,2,3$ refers to the FM injector (F1), the FM detector (F2) and middle metallic wire (M), respectively. $d$ is the distance between the FM injector and the middle wire. The spin resistance of the middle wire is defined as $R_{F3} = R_M = \frac{\rho_M \lambda_M}{w_M w_N \tanh[t_M/\lambda_M]}$. The geometrical factors $w_M$ and $t_M$ are the width and thickness of the middle wire, being $\lambda_M$ its spin diffusion length. When $w_M$ becomes comparable to $\lambda_N$, Supplementary Equation 7 is not accurate anymore, as discussed in ref. 10). However, this is not the case in our samples, because the spin diffusion length of our graphene is much longer ($\lambda_{Gr}$ = 1.20 ± 0.02 μm) than the width of the Pt wire ($w_{Pt}$ = 198 nm) and, thus, the use of the 1D model is valid. Experimentally, this validity has already been proven in devices with Pt spin absorber and Cu spin transport channel[11]. Table 1 in ref. 11 shows that the obtained spin diffusion length and spin Hall angle using the 1D model have a very small deviation with respect to the 3D model, confirming the validity of our approach.

We assume the FM injector and detector spin polarizations are identical ($P_{Co1} = P_{Co2} = P_{Co}$ and $P_{ICo1} = P_{ICo2} = P_{ICo}$), but we consider their different interface resistances. Therefore, we can write for our case:

$r_1 = 2Q_{ICo1} + 2Q_{Co1} + 1$,
$r_2 = 2Q_{ICo2} + 2Q_{Co2} + 1$,
$r_3 = 2Q_{IPt} + 2Q_{Pt} + 1$,
$\hat{\alpha}_1 = P_{ICo} Q_{ICo1} + P_{Co} Q_{Co1}$,
$\hat{\alpha}_2 = P_{ICo} Q_{ICo2} + P_{Co} Q_{Co2}$

where the subscripts ICo1 and ICo2 stand for the Co injector/graphene and Co detector/graphene interface, respectively, and the subscript IPt for the Pt/graphene interface.

Taking into account that $Q_{Co1}, Q_{Co2} \ll Q_{ICo1}, Q_{ICo2}$, that $Q_{IPt}, Q_{Pt} \ll 1$, and considering that $L = 2d$, Supplementary Equation 7 can be simplified to:

$$\Delta R_{\text{NL}}^{\text{abs}} = 8 R_{\text{Gr}} Q_{\text{ICo1}} Q_{\text{ICo2}} P_{\text{ICo}}^2 \frac{(Q_{\text{IPt}}+Q_{\text{Pt}})e^{-\frac{L}{\lambda_{\text{Gr}}}}}{(2Q_{\text{ICo1}}+1)(2Q_{\text{ICo2}}+1)-2(Q_{\text{ICo1}}+Q_{\text{ICo2}}+1)e^{-\frac{L}{\lambda_{\text{Gr}}}}+e^{-\frac{2L}{\lambda_{\text{Gr}}}}}. \quad (8)$$

We fix the following parameters in the above equation: $P_{\text{ICo}} = 0.068$, $\lambda_{\text{Gr}} = 1.20$ μm, $R_{\text{Gr}}^{\square} = 1085$ Ω, $L = 1.27$ μm, $w_{\text{Gr}} = 250$ nm, $w_{\text{Pt}} = 198$ nm, $R_{\text{ICo1}} = 14.7$ kΩ, $R_{\text{ICo2}} = 15$ kΩ, $\rho_{\text{Pt}} = 99$ μΩ·cm (50 K) and 134 μΩ·cm (300 K) and $t_{\text{Pt}} = 21$ nm. We are left with two parameters that are crucial for the spin absorption: $\lambda_{\text{Pt}}$ and $R_{\text{IPt}}$.

We measured directly the interface resistance between Pt and graphene, $R_{\text{IPt}}$ by using a 4-point configuration in the graphene/Pt cross-shaped junction. The measured values are negative, ranging from -8.5 Ω (10 K) to -13 Ω (300 K). This is an artifact which occurs when the resistance of the channel is of the order or higher than the interface resistance due to an inhomogeneous current distribution in this geometry, which is expected due to the large sheet resistance of graphene[12,13]. Its precise value can be determined when combining the results of the spin absorption described by Supplementary Equation 8 with the results of the ISHE experiments described by Supplementary Equation 9 (see Supplementary Note 5).

**Supplementary Note 5: Inverse spin Hall effect by Pt**

The spin-to-charge conversion signal $\Delta R_{\text{SCC}}$ of the ISHE experiment is given by the following expression [7,14,15]:

$$\Delta R_{\text{SCC}} = \frac{2\theta_{\text{SH}} \rho_M x_{M/N}}{w_M} \left( \frac{\overline{I_S}}{I_C} \right), \quad (9)$$

where $\theta_{\text{SH}}$ is the spin Hall angle of the middle wire (M) and $x_{M/N}$ is the correction factor that considers the current in M shunted through the non-magnetic channel (N) (ref. 7). $\overline{I_S}$ is the effective spin current injected vertically into the M wire that contributes to the ISHE, because the spin current at the M/N interface $I_S(z = 0)$ is diluted into the M thickness. To calculate $\overline{I_S}$, we integrate the spin current injected into the M wire and then divide it by the M thickness [7, 14, 15]:

$$\frac{\overline{I_S}}{I_C} \equiv \frac{\int_0^{t_M} I_S(z) dz}{t_M I_C} = \frac{\lambda_M}{t_M} \frac{\left(1-e^{-\frac{t_M}{\lambda_M}}\right)^2}{1-e^{-\frac{2t_M}{\lambda_M}}} \frac{I_S(z=0)}{I_C}, \quad (10)$$

where $I_S(z = 0)$ can be calculated using the following equation:

$$\frac{I_S(z=0)}{I_C} = \frac{2\hat{\alpha}_1 \left[ r_2(1-Q_{I3})e^{-\frac{d}{\lambda_N}} - (1+Q_{I3})e^{-\frac{(2L-d)}{\lambda_N}} \right]}{r_1 r_2 (r_3 - Q_{I3}) - r_1(1+Q_{I3})e^{-\frac{2(L-d)}{\lambda_N}} - r_2(1-Q_{I3})e^{-\frac{2d}{\lambda_N}} - (r_3 - Q_{I3})e^{-\frac{2L}{\lambda_N}} + 2e^{-\frac{2L}{\lambda_N}}}. \quad (11)$$

Taking into account that $Q_{Co1}, Q_{Co2} \ll Q_{ICo1}, Q_{ICo2}$, that $Q_{IPt}, Q_{Pt} \ll 1$, and considering that $L = 2d$, Supplementary Equation 10 can be simplified to:

$$\frac{\bar{I}_S}{I_C} = \frac{\lambda_{Pt}}{t_{Pt}} \frac{\left(1-e^{-\frac{t_{Pt}}{\lambda_{Pt}}}\right)^2}{1-e^{-\frac{2t_{Pt}}{\lambda_{Pt}}}} \frac{2P_{ICo}Q_{ICo1}\left[(2Q_{ICo2}+1)e^{\frac{L}{2\lambda_{Gr}}} - e^{-\frac{L}{2\lambda_{Gr}}}\right]}{(2Q_{ICo1}+1)(2Q_{ICo2}+1)e^{\frac{L}{\lambda_{Gr}}} - 2(Q_{ICo1}+Q_{ICo2}+1) + e^{-\frac{L}{\lambda_{Gr}}}}. \quad (12)$$

In the case of Pt/graphene cross junction, the equivalent sheet resistance of graphene and Pt is 1085 Ω and 64 Ω at 300 K, respectively. Therefore, the shunting coefficient $x_{Pt/Gr}$ is expected to be very close to unity. Using $x_{Pt/Gr} = 1$, $\theta_{SH} = 17.8 \pm 2.0\%$ (50 K) and $23.4 \pm 2.5\%$ (300 K) from ref. 14, and the parameters used for Supplementary Equation S8 (Supplementary Note 4), we are left with the same two parameters: $\lambda_{Pt}$ and $R_{IPt}$.

By using the experimental results from the spin absorption [$\Delta R_{NL}^{abs} = 24.5 \pm 0.9$ mΩ (50 K) and $22 \pm 1$ mΩ (300 K)] and ISHE experiments [$\Delta R_{SCC} = 5.9 \pm 0.2$ mΩ (50 K) and $11.2 \pm 0.7$ mΩ (300 K)], we can solve Supplementary Equations 8 and 9 simultaneously to extract the unknown values for $\lambda_{Pt}$ and $R_{IPt}$. We obtain $\lambda_{Pt} = 2.1 \pm 0.3$ nm and $R_{IPt} = 10.6 \pm 0.4$ Ω at 50 K and $\lambda_{Pt} = 2.1 \pm 0.4$ nm and $R_{IPt} = 8.4 \pm 0.4$ Ω at 300 K.

**Supplementary Note 6: Efficiency of spin injection and detection using Pt wires: Experiments and discussion**

We prepared several samples (see, for instance, Supplementary Fig. 4a) with adjacent Pt electrodes to observe generation and detection of spin currents using SHE and ISHE. Unfortunately, a very small signal (~0.01 mΩ) is expected, due to the conductivity mismatch of the two Pt/graphene contacts (instead of one contact only in the cases of spin detection with ISHE or spin injection with SHE reported in the main text). In order to observe a spin signal from the Ohmic baseline in the non-local measurement, a magnetic field of 7 kOe is rotated in plane. In the *x*-direction, the dephasing of the Hanle precession would cancel the spin signal, while in the *y*-direction (the same as the spin polarization), no Hanle effect would occur. A $\cos^2$ dependence would be expected, with an amplitude corresponding to the spin signal. The noise of the measurement (0.1-0.2 mΩ) is larger than the expected signal and, therefore, cannot be observed (Supplementary Fig. 4b).

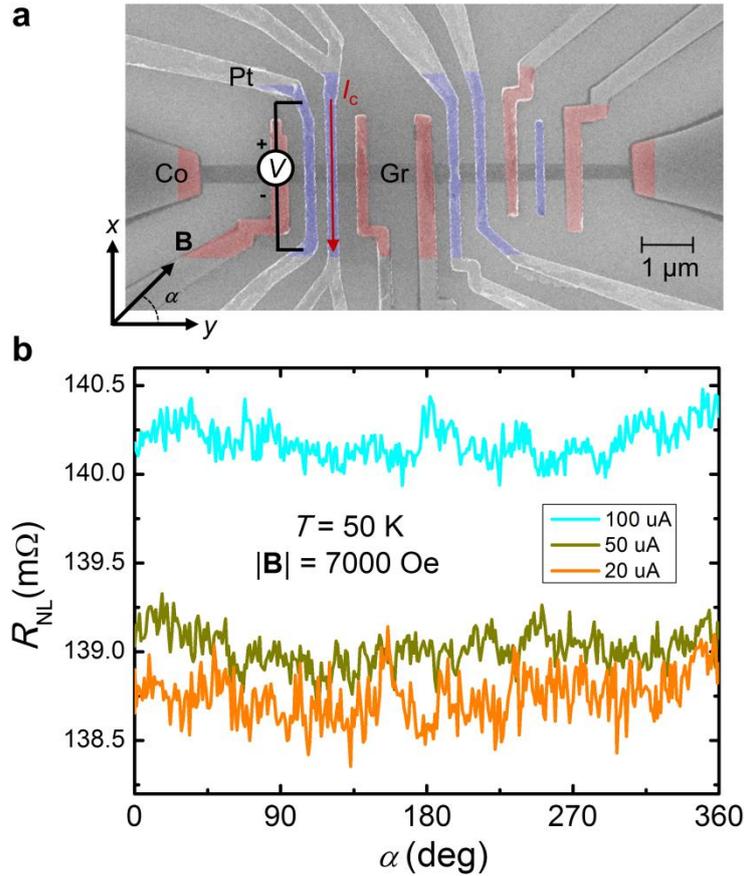

**Supplementary Figure 4 | Simultaneous spin injection and detection in a graphene channel using Pt wires.** (a) SEM image of a graphene/Pt lateral heterostructure with adjacent Pt electrodes (blue color) in a trilayer graphene channel (grey color). Ferromagnetic Co electrodes with TiO$_2$ barrier (red color) placed adjacent to each Pt are used to confirm proper spin injection via SHE or detection via ISHE of the Pt wires following the measurements described in the main text. The measurement configuration shown allows full spin injection and detection using the Pt wires. (b) Non-local resistance as a function of the angle of the applied magnetic field, measured using the configuration shown in (a) at $T$=50 K and $|\mathbf{B}|$=7 kOe with different applied currents.

These results show that a full spin injection and detection with Pt is not useful at this stage due to the low efficiency for spin injection. Nevertheless, in our proof-of-principle device, we showed not only that there is spin injection to graphene using Pt, but also that the overall spin-to-charge conversion of the whole device is more efficient than in conventional lateral spin valves with metallic spin channel. One should be careful when directly comparing the non-local $\Delta R_{\mathrm{NL}}$ signal and the spin-to-charge conversion $\Delta R_{\mathrm{SCC}}$ signal. The former only probes the spin accumulation in the channel (in this case graphene) through a tunnel barrier or high resistive interface leading to a large voltage drop, but it cannot be further utilized, for instance to convert it to charge current for cascading in a spin-based logic circuit or to directly switch a magnetic element via spin transfer torque of the pure spin current. This limitation is equivalent to that observed in the local magnetoresistance of a spin valve: a high resistive interface helps in the spin injection, but is detrimental for the spin detection, because the current cannot flow into the detector [see Fig. 3 in ref. 16]. On the other hand, the configuration of the spin-to-charge conversion consists of a transparent interface through which spins can be absorbed or injected. Here the transport is diffusive and the impedance mismatch plays a role. But the transparent interface is necessary to allow for absorption of the pure spin current in the spin Hall material, which is then converted to a charge current which can be potentially utilized. $\Delta R_{SCC}$ directly probes the charge current generated in the spin Hall metal.

On the application side, the combination of spin injection from one ferromagnetic element where the non-volatile information is stored and subsequent spin-to-charge current conversion in a non-magnetic element is important for cascading in potential applications such as the spin-orbit logic proposed by Intel[17]. Additionally, substituting a FM element by a NM electrode overcomes the necessity of controlling the relative magnetic orientation of a second ferromagnet when used as a detector. For instance, another potential application of our results would be in the spin-based magnetologic device proposed by H. Dery *et al.*, where a graphene spin channel is connected with 5 ferromagnetic electrodes for input, operation and reading out[18,19]. If some of the ferromagnetic electrodes in the magnetologic device can be substituted by a spin Hall metal, this will lead to the control of spin currents by charge current instead of the magnetization of the ferromagnet, as well as to cascading output voltages from one logic element to the next.

**Supplementary References**

1. Idzuchi, H., Fert, A. & Otani, Y. Revisiting the measurement of the spin relaxation time in graphene-based devices. *Phys. Rev. B* **91**, 241407(R) (2015).
2. Valenzuela, S. O. & Tinkham, M. Direct electronic measurement of the spin Hall effect. *Nature* **442**, 176-179 (2006).
3. Mihajlovic, G., Pearson, J. E., Bader, S. D. & Hoffmann A. Surface Spin Flip Probability of Mesoscopic Ag Wires. *Phys. Rev. Lett.* **104**, 237202 (2010).
4. Villamor, E., Isasa, M., Hueso, L. E. & Casanova, F. Temperature dependence of spin polarization in ferromagnetic metals using lateral spin valves. *Phys. Rev. B* **88**, 184411 (2013).
5. Piraux, L., Dubois, S., Fert, A. & Belliard, L. The temperature dependence of the perpendicular giant magnetoresistance in Co/Cu multilayered nanowires. *Eur. Phys. J. B* **4**, 413-420 (1998).
6. Reilly, A. C. *et al.* Giant magnetoresistance of current-perpendicular exchange-biased spin-valves of Co/Cu. *IEEE Trans. Magn.* **34**, 939-941 (1998).
7. Niimi, Y. *et al.* Extrinsic Spin Hall Effect Induced by Iridium Impurities in Copper. *Phys. Rev. Lett.* **106**, 126601 (2011).
8. Niimi, Y. *et al.* Giant Spin Hall Effect Induced by Skew Scattering from Bismuth Impurities inside Thin Film CuBi Alloys. *Phys. Rev. Lett.* **109**, 156602 (2012).
9. Takahashi, S. & Maekawa, S. Spin injection and detection in magnetic nanostructures. *Phys. Rev. B* **67**, 052409 (2003).
10. Laczkowski, P. *et al.* Evaluation of spin diffusion length of AuW alloys using spin absorption experiments in the limit of large spin-orbit interactions. *Phys. Rev. B* **92**, 214405 (2015).
11. Niimi, Y. *et al.* Extrinsic spin Hall effects measured with lateral spin valve structures. *Phys. Rev. B* **89**, 054401 (2014).
12. Pedersen, R. J. & Vernon, F. L. Effect of film resistance on low-impedance tunneling measurements. *Appl. Phys. Lett.* **10**, 29-31 (1967).
13. Pomeroy, J. M. & Grube, H. "Negative resistance" errors in four-point measurements of tunnel junctions and other crossed-wire devices. *J. Appl. Phys.* **105**, 094503 (2009).
14. Sagasta, E. *et al.* Tuning the spin Hall effect of Pt from the moderately dirty to the superclean regime. *Phys. Rev. B* **94**, 060412(R) (2016).
15. Morota, M. *et al.* Indication of intrinsic spin Hall effect in 4d and 5d transition metals. *Phys. Rev. B* **83**, 174405 (2011).
16. Fert, A. & Jaffrès, H. Conditions for efficient spin injection from a ferromagnetic metal into a semiconductor. *Phys. Rev. B* **64**, 184420 (2001).

17. Manipatruni, S., Nikonov, D. E. & Young, I. A. Spin-orbit logic with magnetoelectric nodes: a scalable charge mediated nonvolatile spintronic logic. Preprint at https://arxiv.org/abs/1512.05428 (2017).
18. Dery, H. *et al.* Spin-based logic in semiconductors for reconfigurable large-scale circuits. *Nature* **447**, 573-576 (2007).
19. Dery, H. *et al.* Nanospintronics based on magnetologic gates. *IEEE Trans. Electron Devices* **59**, 259-262 (2012).